# Integrability and linearizability conditions for a cubic Lotka-Volterra systems

[1]Hersh M. Saber

Salahaddin University Department of Physics, College of Education Shaqlawa, University of Salahaddin, Erbil, Kurdistan Region, Iraq.

Hersh.Saber@su.edu.krd

[2]Waleed H. Aziz

Salahaddin University Department of mathematics, College of Science, University of Salahaddin, Erbil, Kurdistan Region, Iraq

Waleed.Aziz@su.edu.krd

## ABSTRACT

The main objective of this work is to investigate the integrability and linearizability problems around a singular point at the origin of the family of differential systems on $\mathbb{C}[x, y, z]$ of the form $x = \lambda x + p(x, y, z)$, $y = \mu y + q(x, y, z)$, $z = \nu z + r(x, y, z)$, where $p, q$ and $r$ are homogeneous polynomials of degree three (either of which may be zero) and $\lambda \mu \nu \neq 0$. Particularly we are interested where $(\lambda : \mu : \nu) = (1: -1: 1)$.

We give a complete set of necessary conditions for both integrability and linearizability problems.

The local existences of two independent first integrals guarantee that these conditions are sufficient as well. Furthermore, we use Darboux method of integrability and some other techniques to show that the system admits two independent first integrals at the origin.

## INTRODUCTION

This work is devoted to studying the local integrability and linearizability problems of the three-dimensional Lotka-Volterra equations

$$\begin{aligned}
\dot{x} &= x(1 + axy + byz + cxz + L_1 x^2 + U_1 y^2 + V_1 z^2) = P(x, y, z), \\
\dot{y} &= y(-1 + dxy + eyz + fxz + L_2 x^2 + U_2 y^2 + V_2 z^2) = Q(x, y, z), \qquad (2.1)\\
\dot{z} &= z(1 + gxy + hyz + kxz + L_3 x^2 + U_3 y^2 + V_3 z^2) = R(x, y, z).
\end{aligned}$$

The classical Lotka-Volterra system, also called the predator-prey system is traced back of the works [44, 51], that involved in many applications on various subjects to model processes or reactions in diverse fields of science, such as biochemical interactions, from genetic circuits [8] to the market economy [50]. Some other applications that modeled by (2.1), for instance analyzing resource allocation on Cloud computing [31, 30], chemical kinetic [47], neural networks [48] and the coupling of waves in laser physics [38], often produce sustained oscillations modeled with Lotka-Volterra systems.

The qualitative theory of dynamical systems is employed to analyze the behavior of these models. See [37, 7] and references therein. In that sense, one of the important features is the existence of first integrals of the differential systems. This is mainly due to the fact that the existence of a first integral allows reducing the dimension of the system by one. So the methods that allow detecting the presence of first integrals are important in the qualitative theory of the differential systems.

In the last years, much research effort has been devoted to investigating of integrability problems for differential systems, see [24, 3, 23, 28] among others. Up to now just for some simple cases first integral form can be calculated because this problem is very hard and no general methods are available. Nevertheless, by using computer algebra facilities several approaches were developed. The most important among them, singularity analysis [29, 9] and the Darboux method [40, 10]. Moreover, authors in [41, 14, 11], considered integrability problems by using Darboux method for this kind of Lotka-Volterra systems and gave conditions of the existence of first integrals.

For two-dimensional quadratic Lotka-Volterra system
$$\dot{x} = x(\lambda + ax + by), \quad \dot{y} = y(\mu + cx + dy),$$
where $\lambda, \mu \in \mathbb{Z}, \lambda\mu < 0$. Some works have been devoted to investigating this problem for instance [18, 39, 32]. For more general quadratic systems the authors in [17, 25, 54], characterize the existence of a local analytic first integral by using a simple generalization of the Poincaré center-focus problem and necessary conditions were given. Where $\lambda + \mu = 0$, the system has a complexified version of the classical center-focus problem. Other works for $(p: -q)$-resonance on integrability and linearizability of Lotka-Volterra systems were considered in [13, 26, 28, 27, 52, 49, 35].

For cubic systems, the integrability and linearizability of some families of systems are studied in [33, 22, 12, 45], despite these authors, others have thought and attempted for working on three dimensional systems to give necessary and sufficient conditions for both integrability and linearizability for instance [34, 11, 10].

Recently [6, 5, 4] investigated local integrability and linearizability of a quadratic three-dimensional Lotka-Volterra and a particular case in general quadratic three-dimensional differential systems.

In this work, we generalize the work in [4] by considering three-dimensional cubic Lotka-Volterra differential systems and give necessary and sufficient conditions for both integrability and linearizability problems at the origin. Despite this fact, we investigate existence and non-existence polynomial first integral of the integrability cases that we found.

## 3.1 Background

In this section, we give some background materials on the integrability and linearizability problems of differential equations. In particular, we introduce some definitions, theorems and known results that will be used in this paper.

Consider a three dimensional polynomial system

$$\begin{aligned} \dot{x} &= \lambda x + p(x,y,z) = P(x,y,z), \\ \dot{y} &= \mu y + q(x,y,z) = Q(x,y,z), \\ \dot{z} &= \nu z + r(x,y,z) = R(x,y,z), \end{aligned} \quad (3.2)$$

where $\lambda\mu\nu \neq 0$ and $p,q,r \in \mathbb{C}[x,y,z]$. We are interested in the case where $(\lambda:\mu)$ and $(\mu:\nu)$ are two independent resonances and eigenvalues are located in the Siegel domain. Therefore, we can assume (after a possible scaling of time) that $\lambda, \mu, \nu \in \mathbb{Z}$ with $\gcd(\lambda,\mu,\nu) = 1$, $\lambda, \nu > 0$ and $\mu < 0$. In this case we say that the origin has $(\lambda:\mu:\nu)$-resonance. additionally, we define the degree of the polynomial differential system as $n = \max\{\deg(P), \deg(Q), \deg(R)\}$ or simply $\deg\mathcal{X} = n$.

Associated to the polynomial differential system (3.2), there is a polynomial vector field

$$\mathcal{X} = P(x,y,z)\frac{\partial}{\partial x} + Q(x,y,z)\frac{\partial}{\partial y} + R(x,y,z)\frac{\partial}{\partial z}. \quad (3.3)$$

**Definition** [[15, P.~71]] The system (3.2) is *integrable* at the origin if and only if there is a change of coordinates

$$(X, Y, Z) = (x + o(x,y,z), y + o(x,y,z), z + o(x,y,z)), \quad (3.4)$$

which transform system (3.2) into

$$\begin{aligned}\dot{X} &= \lambda X \zeta(x, y, z), \\ \dot{Y} &= \mu Y \zeta(x, y, z), \\ \dot{Z} &= \nu Z \zeta(x, y, z),\end{aligned} \qquad (3.5)$$

where $\zeta = 1 + o(x, y, z)$. Then $X^{-\mu}Y^{\lambda}$ and $Y^{\nu}Z^{-\mu}$ are first integrals of system (3.2).

[[21, P.~72]] The system (3.2) is *linearizable* at the origin if and only if there is a change of coordinates (3.4) which transform system (3.2) into

$$\begin{aligned}\dot{X} &= \lambda X, \\ \dot{Y} &= \mu Y, \\ \dot{Z} &= \nu Z.\end{aligned} \qquad (3.6)$$

A node is linearizable if and only if it has two analytic separatrices. For more detail, see [16, P.~14]

## 3.2 First integrals and inverse Jacobi multiplier

In this subsection we recall definition of first integral with its related concept inverse Jacobi multiplier that they play an important role in this work.

**Definition** [[21, P.~214]] Let $U$ be an open subset of $\mathbb{C}[x, y, z]$. A non-constant analytic function $H: U \to \mathbb{C}$ is said to be *first integral* of a vector field $\mathcal{X}$ on $U$ if it is constant on all solutions $(x(t), y(t), z(t))$ of differential system (3.2) in $U$. This means that, $H(x, y, z)$ is first integral if and only if satisfy the following partial differential equation

$$\mathcal{X}H = P(x,y,z)\frac{\partial H}{\partial x} + Q(x,y,z)\frac{\partial H}{\partial y} + R(x,y,z)\frac{\partial H}{\partial z} = 0. \qquad (3.7)$$

**Definition** [[53, P.~45]] Let $U$ be an open subset of $\mathbb{C}[x, y, z]$. A function $M: U \to \mathbb{C}$ is an *inverse Jacobi multiplier* of system (3.2) in $U$ if $M$ verifies the partial differential equation

$$\mathcal{X}(M) = M \operatorname{div}(\mathcal{X}),$$

where $\operatorname{div}(\mathcal{X})$ denotes the divergence of the vector field

$$\operatorname{div}(\mathcal{X}) = \frac{\partial P(x,y,z)}{\partial x} + \frac{\partial Q(x,y,z)}{\partial y} + \frac{\partial R(x,y,z)}{\partial z}. \qquad (3.8)$$

If $n = 2$, the function $M$ is called an inverse integrating factor. Despite the fact, if we have two independent first integrals then this guarantee that there is an inverse Jacobi multiplier. Conversely, if we have one first integral and an inverse Jacobi multiplier, we can find another first integral.

**Theorem [[4, P.~4070]]** suppose that the analytic vector field

$$x(\lambda + \sum_{|I|>0} A_{x_I} X^I)\frac{\partial}{\partial x} + y(\mu + \sum_{|I|>0} A_{y_I} X^I)\frac{\partial}{\partial y} + z(\nu + \sum_{|I|>0} A_{z_I} X^I)\frac{\partial}{\partial z} = 0, \qquad (3.9)$$

has a first integral $\Phi = x^\alpha y^\beta z^\gamma (1 + O(x,y,z))$, with at least one of $\alpha, \beta, \gamma \neq 0$ and a inverse Jacobi multiplier $M = x^r y^s z^t (1 + O(x,y,z))$ and suppose that the cross product of $(r-i-1, s-j-1, t-k-1)$ and $(\alpha, \beta, \gamma)$ is bounded away from zero for any integer $i,j,k \geq 0$. then the $X$ has a second analytic first integral of the form $\Psi = x^{1-r} y^{1-s} z^{1-t}(1 + O(x,y,z))$, hence the system (3.2) is integrable.

### 3.3 Darboux theory of integrability

Darboux theorem [19] is a method that can be used to construct first integrals and inverse Jacobi multipliers for planar and higher differential systems if the system possess an adequate number of invariant algebraic curves and exponential factors. In this sense, the first integrals and inverse Jacobi multipliers can express as a product of these curves with exponential factors.

**Theorem [[53, P.~93]]** Assume that the polynomial differential system (3.2) has $p$ distinct invariant algebraic surfaces $\ell_i = 0$ with the associated cofactors $L_{f_i}$ for $i = 1, \ldots, p$ and $q$ independent exponential factors $E_j$ with the associated cofactors $L_{E_j}$ for $j = 1, \ldots, q$. Then the following statements hold.

1. The function

$$M = \ell_1^{\lambda_1} \ldots \ell_p^{\lambda_p} E_1^{\mu_1} \ldots E_q^{\mu_q},$$

is an inverse Jacobi multiplier provided that the condition

$$\sum_{i=1}^{p} \lambda_i L_{\ell_i} + \sum_{j=1}^{q} \mu_j L_{E_j} = \text{div}\mathcal{X},$$

is satisfies for certain complex numbers $\lambda_i, i = 1, \ldots, p$ and $\mu_j, j = 1, \ldots, q$.

2. There exists $\lambda_i, \mu_j \in \mathbb{C}$ not all zero such that

$$\sum_{i=1}^{p} \lambda_i L_{\ell_i} + \sum_{j=1}^{q} \mu_j L_{E_j} = 0,$$

if and only if the (multi-valued) function

$$\Phi = \ell_1^{\lambda_1} \ldots \ell_p^{\lambda_p} E_1^{\mu_1} \ldots E_q^{\mu_q},$$

is a first integral of the vector field $\mathcal{X}$.

**Definition** [[15, P.~17]] Let $\ell \in \mathbb{C}[x, y, z]$ be a polynomial. If the algebraic surface $\ell = 0$ is invariant by a vector field $\mathcal{X}$ of degree $n$, then $L_\ell$ is a polynomial of degree at most $n - 1$. In this case we say that $\ell = 0$ is *invariant algebraic surface* of $\mathcal{X}$ and $L_\ell$ is its cofactor. This means that

$$\dot{\ell} = \mathcal{X}\ell = P\frac{\partial \ell}{\partial x} + Q\frac{\partial \ell}{\partial y} + R\frac{\partial \ell}{\partial z} = L_\ell \ell, \tag{3.10}$$

for some polynomial $L_\ell \in \mathbb{C}[x, y, z]$.

**Definition** [[15, P.~19]] Let $p$ be a singular point of the vector field $\mathcal{X}$. Then if $\ell$ is an invariant algebraic curve of $\mathcal{X}$ which does not vanish at p, its cofactor $L_\ell$ must vanish at $p$. Furthermore, if $E = \exp(g/\ell)$ is an exponential factor of $\mathcal{X}$, then $L_E$ must vanish at $p$ too. For general case an invariant algebraic surface can be found by using the concept of undetermined coefficients. Moreover, we seek for a polynomial of degree $m$ that they go through origin or not such as if $\zeta = 1, 0$ then

$$\ell = \zeta + a_{[1,0,0]}x + a_{[0,1,0]}y + a_{[0,0,1]}z + \cdots + a_{[0,0,m]}z^m,$$

with cofactor degree $n - 1$

$$\ell = 1 - \zeta + b_{[1,0,0]}x + b_{[0,1,0]}y + b_{[0,0,1]}z + \cdots + a_{[0,0,n-1]}z^{n-1},$$

where $n$ is degree of a system. For more detail, see [43] among others.

**Theorem** [[42, P.~445]] Suppose $\ell \in \mathbb{C}[x, y, z]$ and let $\ell = \ell_1^{n_1}, \ldots, \ell_r^{n_r}$ be its factorization in irreducible factors over $\mathbb{C}[x, y, z]$. Then for a vector field $\mathcal{X}$, $\ell = 0$ is an invariant algebraic surface with cofactor $L_\ell$ if and only if $\ell_i = 0$ is an invariant algebraic surface for each $i = 1, \ldots, r$ with cofactor $L_{\ell_i}$. Moreover $L_\ell = n_1 L_{\ell_1} + \cdots + n_r L_{\ell_r}$. Suppose $f$ and $g$ are invariant algebraic surfaces with respective cofactors $L_f$ and $L_g$. Then

1. If $L_f = L_g$ then $f + g$ is an invariant algebraic surface with cofactor $L_f = L_g$.
2. $f^k$ for $k \in \mathbb{C}$ is an invariant algebraic surface with cofactor $kL_f$.

Proof: Is clear.

**Definition** [[42, P.~446]] An exponential function $E$ of the form $E(x, y, z) = \exp(f(x, y, z)/g(x, y, z))$ where $f, g \in \mathbb{C}[x, y, z]$, is called *exponential factor* if satisfy the following equation,

$$\mathcal{X}E = L_E E, \tag{3.11}$$

where $L_E$ is a cofactor of $E$ and has one degree less than the vector field $\mathcal{X}$.

**Theorem** [[42, P.~448]] Let $E = \exp(f/g)$ be an exponential factor with the cofactor $L_E$ of a vector field $\mathcal{X}$, then $f = 0$ is an invariant algebraic surface with cofactor $L_f$ and $g$ satisfies the equation

$$\mathcal{X}(g) = gL_f + fL_E.$$

**Definition** [[15, P.~18]] A (multi-valued) function is said to be *Darboux function* if it is of the form

$$E^{g/h} \prod_{i=1}^{r} F_i^{l_i}$$

where the $F_i, g$ and $h$ are polynomials, and $l_i$ are complex numbers.

### 3.4 Normal Forms

Normal form theory is a classical tool provides one of the most powerful techniques in the analysis of nonlinear dynamical systems. The main idea is to simplify equations of a nonlinear dynamical systems by consists of employing successive near identity nonlinear transformations that lead to a differential equation in a simpler form as far as possible, qualitatively equivalent to the original system in the near of a singular point. This theory can be traced back to Poincaré were the first to bring forth the theory in a more definite form.

Consider $n$-dimensional differential system

$$\dot{x} = Ax + \sum_{i=2}^{\infty} f_i(x), \quad x \in (\mathbb{R}^n, 0), \tag{3.12}$$

the $f_i$'s are vector-valued homogeneous polynomials of degree $i$ in $(\mathbb{R}^n,0)$, where $(\mathbb{R}^n,0)$ denotes a neighborhood of the origin in $\mathbb{R}^n$. If the series in (3.12) is convergent, then the system is analytic.

Let $A$ be an $n \times n$ matrix with eigenvalues $\alpha_1, \ldots, \alpha_n$.

1. The eigenvalues of $A$ satisfy a resonant relation if there exists some $j \in \{1, \ldots, n\}$ and $k = (k_1, \ldots, k_n) \in \mathbb{Z}_+^n$, such that

$$\alpha_i = \langle k, \alpha \rangle = \sum_{i=1}^{n} k_i \alpha_i, \quad 15pt|k|2, \tag{3.13}$$

where $\mathbb{Z}_+ = \mathbb{N} \cup \{0\}$ and $|k| = \sum_{i=1}^{n} k_i$ is the degree of a monomial.

2. Otherwise the eigenvalues of $A$ do not satisfy any resonant relation.

3. The coefficient $x_i^k$ of the monomial $x$ is called a resonant coefficient and the corresponding term is called a resonant term. See for more detail [53, 36] and the references therein.

**Theorem** [(*Poincaré linearization theorem*) [53, P.~56]] If the eigenvalues of $A$ in (3.12) do not satisfy any resonant relation, then system (3.12) is linearizable.

**Theorem** [(*Poincaré Dulac normal form theorem*) [53, P.~57]] An analytic or a formal differential system (3.12) is always formally equivalent to its Poincaré normal form via a near identity formal transformation.

### 3.4.1 Homological Operator

A space $\mathbb{H}_k$ denote the vector space of the homogeneous polynomials defined as follows
$$\mathbb{H}_k = \{\text{homogeneous polynomials of degree } k \text{ in } \quad 5pt x \in \mathbb{R}^n\}.$$

Clearly, the set of monomials $x^k = x_1^{k_1} x_2^{k_2} \ldots x_n^{k_n}$ is a basis for the vector space $\mathbb{H}_k$.

The homological operator is a linear operator $\mathcal{L}: \mathbb{H}_k^n \to \mathbb{H}_k^n$ is defined by
$$\mathcal{L}h(y) = Dh(y)Ay - Ah(y).$$

By performing a sequence of transformations of the form
$$x = H(y) = y + h^k(y), \quad 15pt h^k \in \mathbb{H}_k.$$

we can remove all non-resonant terms of system (3.12) and is formally equivalent to
$$\dot{y} = Ay + G(y).$$

For more detail see [46, P.~282].

**Definition** [[36, P.~48]] The *Poincaré domain* $\mathcal{B} \subset \mathbb{C}^n$ is the collection of all tuples $\alpha = (\alpha_1, \ldots, \alpha_n)$ such that the convex hull of the point set $\{\alpha_1, \ldots, \alpha_n\}$ does not contain the origin inside or on the boundary.

**Definition** [[36, P.~48]] The *Siegel domain* $\mathcal{G}$ is the complement of the Poincaré domain in $\mathbb{C}^n$.

**Theorem** [(*Poincaré normal form theorem*) [53, P.~294]] If the eigenvalues of the linear part $A$ of the analytic differential system (3.12) belong to the Poincaré domain, then system (3.12) is analytically equivalent to a polynomial differential system.

**Theorem** [[53, P.~294]] If the eigenvalues $\alpha = (\alpha_1, \ldots, \alpha_n)$ of $A$ belong to the Poincaré domain, then $\alpha$ satisfy finitely many resonant relations, that is $\alpha_i$ in equation

(3.13) contains finitely many elements.

### 3.4.2 Reduction to the Poincaré domain

Let $\alpha$ be a vector of eigenvalues in the Poincaré domain then one can obtain linearization results when there are resonances that is allowed to eliminate non-resonant terms by a formal change of variables. We use this principle in two ways. Firstly, in many cases we can choose a coordinate system so that two of the variables decouple to give a linearizable node at the origin. If this is so, it just remains to find a linearizing transformation for the third variable via some simple power series arguments. Secondly, and more rarely, we can perform the Theorem 3. Since the new system is linearizable, thus we can find two first integrals which we can pull back to first integrals of the original system.

**Theorem** [[4, P.~4070]] If the system (2.1) is integrable and there exist a function $\zeta = x^\alpha y^\beta z^\gamma (1 + O(x, y, z))$ such that $\mathcal{X}(\zeta) = k\zeta$ for some constant $k = \alpha\lambda + \beta\mu + \gamma\nu$, then the system is linearizable.

### 4.1 Three dimensional Lotka-Volterra system

In this section, we begin by illustrate and explain the mechanism of how one can obtain necessary and sufficient conditions at the origin for integrability and linearizability of the system (2.1). On one hand we drive a collection of necessary conditions for integrability and linearizability that guarantee system (2.1) admits two analytic first integrals. On the other hand for proving their sufficiency, we show that the system is completely integrable at the origin. In particular, we use Darboux theory of integrability, linearizable node and transformation technique to show that the system admits two independent first integral.

Consider three dimensional Lotka-Volterra system
$$\begin{aligned} \dot{x} &= x(\lambda + axy + byz + cxz + L_1 x^2 + U_1 y^2 + V_1 z^2) = P(x, y, z), \\ \dot{y} &= y(\mu + dxy + eyz + fxz + L_2 x^2 + U_2 y^2 + V_2 z^2) = Q(x, y, z), \\ \dot{z} &= z(\nu + gxy + hyz + kxz + L_3 x^2 + U_3 y^3 + V_3 z^2) = R(x, y, z), \end{aligned} \qquad (4.14)$$
where $\lambda, \mu, \nu \neq 0$. According to Definition 3.2, system (4.14) has two independent first integrals of the form
$$H_1 = x^{-\mu} y^\lambda (1 + O(x, y, z)), \quad \text{and} \quad H_2 = y^\nu z^{-\mu} (1 + O(x, y, z)) \qquad (4.15)$$

## 4.2 Mechanisms for integrability and linearizability

In this section, we present the mechanism that will be used for finding the integrability conditions for the system (4.14).

Let $H_1$ and $H_2$ be two independent first integrals of the form

$$H_1 = xy(1 + O(x,y,z)), \quad \text{and} \quad H_2 = yz(1 + O(x,y,z)), \tag{4.16}$$

such that

$$\dot{H}_1 = \sum_{n_1,n_2 \geq 0} \xi_{n_1,n_2} x^{n_1} y^{n_1+n_2} z^{n_2},$$

and

$$\dot{H}_2 = \sum_{n_1,n_2 \geq 0} \kappa_{n_1,n_2} x^{n_1} y^{n_1+n_2} z^{n_2},$$

where $\lambda n_1 + \mu(n_1 + n_2) + \nu n_2 = 0$, for all $n_1, n_2 \in \mathbb{N}$. The coefficients $\xi_{n_1,n_2}$ and $\kappa_{n_1,n_2}$ are polynomials in parameters in (4.14) which are obstructions of the existence of such first integrals $H_1$ and $H_2$. We denote by I=$\langle \xi_{n_1,n_2}, \kappa_{n_1,n_2} \rangle$ the ideal generated by such polynomials $\xi_{n_1,n_2}$ and $\kappa_{n_1,n_2}$ and its variety by $V(I)$. Moreover, we calculate such polynomials to a given degree using Maple. The degree was used 16.

Hence we calculate the irreducible decomposition of the variety $V$ by computing a factorized Gröbner basis using the Routine mianAssGTZ in Singular [20]. In particular, the conditions on parameters that make $\xi_{n_1,n_2}$ and $\kappa_{n_1,n_2}$ vanish. This will be correspond to necessary conditions for integrability of (4.14) at the origin.

For linearizability, we proceeded similarly: computing the conditions for the existence of a linearizing change of coordinate up to some finite order to find necessary conditions, and exhibiting a linearizing change of coordinates for sufficiency. In this case, the first integrals can be obtained easily by pulling back the first integral of the linearized system (4.14).

## 4.3 integrability and linearizability conditions

The Lotka-Volterra system (4.14) has a property that $x = 0, y = 0, z = 0$ are always invariant algebraic surfaces with respective cofactors

$$L_x = \lambda + ax + by + cz + L_1 x^2 + V_1 z^2,$$
$$L_y = \mu + dx + ey + fz + L_2 x^2 + V_2 z^2,$$
$$L_z = \nu + gx + hy + kz + L_3 x^2 + V_3 z^2.$$

We now are looking for two independent analytic first integrals

$$H_1 = xy(1 + O(x,y,z)), \text{ and } H_2 = yz(1 + O(x,y,z)).$$

For finding resonant terms we written $H_1$ and $H_2$ as power series of order 16 then we calculate the obstruction terms for the existences two analytic first integrals. We denote them by $\xi_{n,n}$ and $\kappa_{n,n}$ for $n = 1,...,5$ then

$$X(H_1) = P\frac{\partial H_1}{\partial x} + Q\frac{\partial H_1}{\partial y} + R\frac{\partial H_1}{\partial z} = \sum_{\substack{j=2,3 \\ i=0,...,j}} \xi_{k,k} x^{j-i} y^j z^i, \quad k = 1,...,5,$$

and

$$X(H_2) = P\frac{\partial H_2}{\partial x} + Q\frac{\partial H_2}{\partial y} + R\frac{\partial H_2}{\partial z} = \sum_{\substack{j=2,3 \\ i=0,...,j}} \kappa_{k,k} x^{j-i} y^j z^i, \quad k = 1,...,5.$$

The expressions for $\xi_{n,n}$ and $\kappa_{n,n}$ for $n = 1,...,5$ are presented and will no present the others as the size of these polynomials are dramatically increases.

$\xi_{1,1} = a + d.$

$\xi_{2,2} = b + e.$

$\xi_{3,3} = -L_1 U_1 + L_2 U_2.$

$\xi_{4,4} = U_1 V_2 + U_2 V_2 - U_3 V_1 - U_3 V_2.$

$\xi_{5,5} = -cU_1 - cU_3 + fU_1 + 2fU_2 - fU_3.$

$\kappa_{1,1} = d + g.$

$\kappa_{2,2} = e + h.$

$\kappa_{3,3} = U_2 V_2 - U_3 V_3.$

$\kappa_{4,4} = -L_2 U_1 + L_2 U_2 + L_2 U_3 - L_3 U_1.$

$\kappa_{5,5} = -fU_1 + 2fU_2 + fU_3 - kU_1 - kU_3.$

The integrability conditions then can be summarized in the theorem below.

**Theorem**

System (2.1) is integrable at the origin with $(1:-1:1)$-resonance if and only if one of the following conditions holds

1) $a + d = b + e = d + g = e + h = f = g = h = L_2 = V_2 = U_1 = U_3 = 0.$

2) $a + d = b + e = c = d + g = e + h = f = g = h = k = L_1 = L_2 = L_3 = V_2 = U_3 = 0.$

2*) $a + d = b + e = c = d + g = e + h = f = g = h = k = L_2 = V_1 = V_2 = V_3 = U_1 = 0.$

3) $a + d = b + e = c - 2f = d + g = e + h = g = h = k = L_1 = L_2 = L_3 = V_2 = U_1 - 2U_2 = U_3 = 0$.

3*) $a + d = b + e = c = d + g = e + h = 2f - k = g = h = L_2 = V_1 = V_2 = V_3 = U_1 = 2U_2 - U_3 = 0$.

4) $a + d = b + e = d + g = e + h = U_1 = U_2 = U_3 = 0$.

5) $a + d = b + e = c = d + g = e + h = f = g = h = k = L_1 - L_2 = L_3 = V_2 = U_1 - U_2 = U_3 = 0$.

5*) $a + d = b + e = c = d + g = e + h = f = g = h = k = L_2 = V_1 = V_2 - V_3 = U_1 = U_2 - U_3 = 0$.

6) $a + d = b + e = d + g = e + h = f = g = h = L_1 = L_2 = L_3 = V_1 + V_3 = V_2 - V_3 = U_1 + U_3 = U_2 - U_3 = 0$.

6*) $a + d = b + e = d + g = e + h = f = g = h = L_1 + L_3 = L_2 + L_3 = V_1 = V_2 = V_3 = U_1 + U_3 = U_2 + U_3 = 0$.

7) $a + d = b + e = d + g = e + h = f = g = h = L_1 = L_2 = L_3 = V_1 = V_2 = V_3 = U_1 + U_3 = 0$.

8) $a + d = b + e = d + g = e + h = L_1 U_1 - L_2 U_2 = U_2 V_2 - U_3 V_3 = L_1 V_1 - 2L_1 V_3 + L_3 V_3 = cL_1 + cL_3 - 3kL_1 + kL_3 = cV_1 - 3cV_3 + kV_1 + kV_3 = L_1 U_2 + L_1 U_3 - L_2 U_2 - L_3 U_2 = L_1 V_2 + L_1 V_3 - L_2 V_3 - L_3 V_3 = L_2 U_1 - L_2 U_2 - L_2 U_3 + L_3 U_1 = L_2 V_1 - 2L_2 V_3 + L_3 V_1 + L_3 V_2 - L_3 V_3 = -U_1 V_3 + U_2 V_1 - U_2 V_3 + U_3 V_3 = -U_1 V_2 + U_3 V_1 + U_3 V_2 - U_3 V_3 = cL_2 + cL_3 + 2fL_3 - 3kL_2 - kL_3 = cU_2 + cU_3 - 2fU_2 - kU_2 + kU_3 = cV_2 + cV_3 - 2fV_2 + kV_2 - kV_3 = cU_1 + cU_3 - 4fU_3 + kU_1 - kU_3 = fU_1 - 2fU_2 - fU_3 + kU_1 + kU_3 = fV_1 - 3fV_3 + kV_1 + 2kV_2 - kV_3 = fL_2 U_2 + fL_3 U_1 + kL_1 U_1 - 2kL_2 U_1 - kL_3 U_1 = fL_2 V_3 + fL_3 V_1 + fL_3 V_2 - fL_3 V_2 - fL_3 V_3 - kL_2 V_1 - 2kL_2 V_2 + kL_2 V_3 = fL_1 + fL_3 + kL_1 - 2kL_2 - kL_3$.

Furthermore, the origin of system (1) is linearizable if and only if it satisfies one of the conditions ((1) – (7) except condition (4)) or one of the following holds:

8.1) $a + d = b + e = d + g = e + h = g = h = U_1 = U_2 = U_3 = 0$.

8.2) $a + d = b + e = c - k = d + g = e + h = f - k = g = h = L_1 - L_3 = L_2 - $

$L_3 = V_1 - V_3 = V_2 - V_3 = U_1 - U_3 = U_2 - U_3 = 0.$

8.3) $a + d = b + e = c = d + g = e + h = f = g = h = k = L_1 = L_2 = L_3 = V_2 -$

$V_3 = U_2 - U_3 = V_1 U_3 - V_3 U_1 = 0.$

8.3*) $a + d = b + e = c = d + g = e + h = f = g = h = k = V_1 = V_2 = V_3 = L_1 -$

$L_2 = U_1 - U_2 = L_2 U_3 - L_3 U_2 = 0.$

8.4) $a + d = b + e = c - 2f + k = d + g = e + h = g = h = L_1 = L_2 = L_3 = V_1 =$

$V_2 = V_3 = U_1 - 2U_2 + U_3 = fU_3 - kU_2 = 0.$

8.5) $a + d = b + e = c = d + g = e + h = f = g = h = k = L_1 = L_2 = L_3 = V_1 =$

$V_2 = V_3 = 0.$

**Proof:**

Since 2*, 3*, 5*, 6* and 8.3* are dual with 3, 4, 5, 6 and 8.3 respectively under the transformation $(x, y, z) \to (z, y, x)$, then we do not consider them separately. The other cases are considered below.

**Case 1.** In this case we have

$$\begin{aligned} \dot{x} &= x(1 + cxz + L_1 x^2 + V_1 z^2), \\ \dot{y} &= y(-1 + U_2 y^2), \\ \dot{z} &= z(1 + kxz + L_3 x^2 + V_3 z^2). \end{aligned} \quad (4.17)$$

The subsystem

$$\dot{x} = x(1 + cxz + L_1 x^2 + V_1 z^2),$$
$$\dot{z} = z(1 + kxz + L_3 x^2 + V_3 z^2),$$

is independent in $y$, then they are linearizable node and the linearizing change is given

$$X = x(1 + O(x, z)), \quad \text{and} \quad Z = z(1 + O(x, z)),$$

such that

$$\dot{X} = X, \quad \text{and} \quad \dot{Z} = Z.$$

For the second equation

$$\dot{y} = y(-1 + U_2 y^2). \quad (4.18)$$

We have an invariant algebraic surface with its cofactor

$$\ell = 1 - U_2 y^2 = 0, \quad L_\ell = 2U_2 y^2. \tag{4.19}$$

Then the transformation $Y = y\ell^{-\frac{1}{2}}$, can linearize equation (4.18) and gives $\dot{Y} = -Y$, because $\dot{Y} = -\frac{1}{2}\dot{\ell} y \ell^{-\frac{1}{2}-1} + \dot{y}\ell^{-\frac{1}{2}}$ and using equations (4.18) and (4.19) it is easy to see $\dot{Y} = -Y$.

Consequently, we can obtain two first integrals $XY$ and $YZ$. Therefore, the required first integrals for the original variables are

$$H_1 = xy(1 + O(x,y,z)), \text{ and } H_2 = yz(1 + O(x,y,z)).$$

**Case 2.** System (2.1) reduces to

$$\begin{aligned}\dot{x} &= x(1 + U_1 y^2 + V_1 z^2),\\ \dot{y} &= y(-1 + U_2 y^2),\\ \dot{z} &= z(1 + V_3 z^2).\end{aligned} \tag{4.20}$$

In this case we have invariant algebraic surfaces $\ell_1 = 1 - U_2 y^2 = 0$ and $\ell_2 = 1 + V_3 z^2 = 0$ with cofactors $L_{\ell_2} = 2 U_2 y^2$ and $L_{\ell_2} = 2V_3 z^2$ respectively. These allow us to find first integrals

$$H_1 = xy\ell_1^{-\frac{U_1+U_2}{2U_2}} \ell_2^{-\frac{V_1}{2V_3}}, \text{ and } H_2 = yz\ell_1^{-\frac{1}{2}}\ell_2^{-\frac{1}{2}},$$

Since $\zeta = z(\ell_2)^{-\frac{1}{2}}$ satisfies $\dot{\zeta} = \zeta$, then the system (4.20) must be linearizable by Theorem 3.4.2.

For the case $U_2 = 0$ and $V_3 \neq 0$, also $\ell_2 = 1 + V_3 z^2$ is invariant algebraic surface with cofactor $L_{\ell_2} = 2V_3 z^2$ and an exponential factor $E = e^{\frac{1}{2}U_1 y^2}$ with cofactor $L_E = -U_1 y^2$. One can easily find a first integral $\phi = yz\ell_2^{-\frac{1}{2}}$, and an inverse Jacobi multiplier $M = x^{\frac{V_1-2V_3}{V_1}} y^{\frac{V_1-2V_3}{V_1}} z \ell^2 E^{-\frac{2V_3}{V_1}}$ then using Theorem 3.2 we can find second first integral of the form $\Psi = x^{\frac{2V_3}{V_1}} y^{\frac{2V_3}{V_1}}(1 + O(x,y,z))$. Consequently, it is easy to see

$$H_1 = \Psi^{\frac{V_1}{2V_3}} = xy(1 + O(x,y,z)), \text{ and } H_2 = yz(1 + O(x,y,z)),$$

and it is also linearizable. Moreover, Since $\zeta = y$ satisfies $\dot{\zeta} = \zeta$, then the reduce system with $U_2 = 0$ and $V_3 \neq 0$ must be linearizable by Theorem 3.4.2.

For the cases $V_3 = 0$ and $U_2 \neq 0$, or $V_3 = 0$ and $U_2 = 0$, we can follow the same proceed as befor.

**Case 3.** We obtain the system

$$\begin{aligned}\dot{x} &= x(1 + 2fxz + 2U_2 y^2 + V_1 z^2),\\ \dot{y} &= y(-1 + fxz + U_2 y^2),\\ \dot{z} &= z(1 + V_3 z^2).\end{aligned} \quad (4.27)$$

Note that $\ell = 1 + V_3 z^2 = 0$, is invariant algebraic surface with cofactor $L_\ell = 2V_3 z^2$, that gives an first integral and inverse Jacobi multiplier

$$\Phi = xy^{-2}z^{-3}\ell^{\frac{V_1 - 3V_3}{V_3}}, \quad \text{and} \quad M = xy^3 z^3.$$

Theorem 3.2, guarantee that there is another first integral of the form

$$\Psi = y^{-2}z^{-2}(1 + O(x, y, z)).$$

Then, the desired first integrals are

$$H_1 = \Phi \Psi^{-\frac{3}{2}} = xy(1 + O(x, y, z)), \text{ and } H_2 = \Psi^{-\frac{1}{2}} = yz(1 + O(x, y, z)).$$

Since $\zeta = z\ell^{-\frac{1}{2}}$ satisfies $\dot{\zeta} = \zeta$, then the system (4.27) must be linearizable by Theorem 3.4.2.

If $V_3 = 0$, then we replace the invariant algebraic surface with an exponential factor $E = e^{-\frac{1}{2}V_1 z^2}$ with cofactor $L_E = -V_1 z^2$. Therefore, the reduce system has an first integral and inverse Jacobi multiplier

$$\Phi = xy^{-2}z^{-3}E, \quad \text{and} \quad M = x^3 y^{-1} z^{-3} E.$$

Thus, the second first integral is $\Psi = x^{-2}y^2 z^4(1 + O(x, y, z))$. Hence the first integrals of desired form are

$$H_1 = \Phi^{-2}\Psi^{-\frac{3}{2}} = xy(1 + O(x, y, z)), \text{ and } H_2 = \Phi^{-1}\Psi^{-\frac{1}{2}} = yz(1 + O(x, y, z)).$$

Since the third equation is linearize $\zeta = z$, hence the reduce system is also linearizable.

**Case 4.** The corresponding system of (2.1) is

$$\begin{aligned}\dot{x} &= x(1 + gxy + hyz + cxz + L_1 x^2 + V_1 z^2),\\ \dot{y} &= y(-1 - gxy - hyz + fxz + L_2 x^2 + V_2 z^2),\\ \dot{z} &= z(1 + gxy + hyz + kxz + L_3 x^2 + V_3 z^2).\end{aligned} \quad (4.28)$$

**Case 5.** By substitution integrability conditions, system (2.1) reduce to

$$\begin{aligned}\dot{x} &= x(1 + L_2 x^2 + U_2 y^2 + V_1 z^2), \\ \dot{y} &= y(-1 + L_2 x^2 + U_2 y^2), \\ \dot{z} &= z(1 + V_3 z^2).\end{aligned} \qquad (4.29)$$

There is an invariant algebraic surface of the form $\ell = 1 + V_3 z^2 = 0$ with cofactor $L_\ell = 2 V_3 z^2$. One can find a first integral and an inverse Jacobi multiplier of the form

$$\Phi = x\, y^{-1} z^{-2} \ell^{-\frac{1}{2}(\frac{V_1 - 2V_3}{V_3})}, \quad \text{and} \quad M = x^3 y z^{-1} \ell^{(\frac{-V_1 + 2V_3}{V_3})}.$$

By Theorem 3.2, we can find a second first integral of the form

$$\Psi = x^{-2} z^2 (1 + O(x, y, z)).$$

Therefore, the desired first integrals are of the form

$$H_1 = \Phi^{-1} \Psi^{-1} = xy(1 + O(x,y,z)), \quad \text{and} \quad H_2 = \Phi^{-1} \Psi^{-\frac{1}{2}} = yz(1 + O(x,y,z)).$$

Since $\zeta = z \ell^{-\frac{1}{2}}$ satisfies $\dot{\zeta} = \zeta$, then the system (4.28) must be linearizable by Theorem 3.4.2.

If $V_3 = 0$, then we replace the invariant algebraic surface with an exponential factor $E = e^{-\frac{1}{2} V_1 z^2}$ with cofactor $L_E = -V_1 z^2$. Therefore, the reduce system has an first integral and inverse Jacobi multiplier

$$\Phi = xy^{-1} z^{-2} E, \quad \text{and} \quad M = x^2 y^2 z\, E.$$

Thus, the second first integral is $\Psi = x^{-1} y^{-1}(1 + O(x, y, z))$. Hence the first integrals of desired form are

$$H_1 = \Psi^{-1} = xy(1 + O(x,y,z)), \quad \text{and} \quad H_2 = \Phi^{-\frac{1}{2}} \Psi^{-\frac{1}{2}} = yz(1 + O(x,y,z)).$$

Since the third equation is linearize as $\zeta = z$, hence the reduce system is also linearizable.

**Case 6.** In this case the system is

$$\begin{aligned}\dot{x} &= x(1 + cxz - U_3 y^2 - V_3 z^2), \\ \dot{y} &= y(-1 + U_3 y^2 + V_3 z^2), \\ \dot{z} &= z(1 + kxz + U_3 y^2 + V_3 z^2).\end{aligned} \qquad (4.32)$$

In this case, there is an invariant algebraic surface $\ell = 1 + \frac{1}{2}(c+k)xz = 0$ with cofactor $L_\ell = (c+k)xz$. Therefore, the system has a first integral and an inverse Jacobi multiplier of the form

$$\phi = x^{-\frac{c+k}{c}} y^{-\frac{c+k}{c}} \ell, \quad \text{and} \quad M = x^2 y^3 z^2.$$

Theorem 3.2, guarantee that we can find second first integral of the form

$$\Psi = x^{-1} y^{-2} z^{-1}.$$

Hence the first integrals of the required form are

$$H_1 = \phi^{-\frac{c}{c+k}} = xy(1 + O(x,y,z)), \text{ and } H_2 = \Phi^{-\frac{c}{c+k}} \Psi^{-1} = yz(1 + O(x,y,z)).$$

Since $\zeta = xz\ell^{-1}$ satisfies $\dot\zeta = 2\zeta$, then the system (4.28) must be linearizable by Theorem 3.4.2.

If $c = -k$, then we can find an exponential factor $E = e^{\frac{1}{2}kxz}$ with cofactor $L_E = kxz$. Thus, there is a first integral and an inverse Jacobi multiplier of the form

$$\phi = x\, y\, E, \quad \text{and} \quad M = x^2 y^3 z^2.$$

Theorem 3.2, guarantee that we can find second first integral of the form

$$\Psi = x^{-1} y^{-2} z^{-1}.$$

Hence the first integrals of the required form are

$$H_1 = \phi = xy(1 + O(x,y,z)), \text{ and } H_2 = \Phi^{-1} \Psi^{-1} = yz(1 + O(x,y,z)).$$

Since $\zeta = xz$ satisfies $\dot\zeta = 2\zeta$, then the reduce system must be linearizable by Theorem 3.4.2.

**Case 7.** The subsystem in this case is

$$\begin{aligned} \dot x &= x(1 + cxz - U_3 y^2), \\ \dot y &= y(-1 + U_2 y^2), \\ \dot z &= z(1 + kxz + U_3 y^2). \end{aligned} \quad (4.37)$$

In this case, there is an invariant algebraic surface $\ell = 1 + \frac{1}{2}(c+k)xz = 0$ with cofactor $L_\ell = (c+k)xz$. Therefore, the system has a first integral and an inverse Jacobi multiplier of the form

$$\phi = x^{\frac{1}{2}\frac{U_2+U_3}{U_3}} y z^{-\frac{1}{2}\frac{U_2-U_3}{U_3}} \ell^{-\frac{1}{2}\frac{(c+k)U_3+(c-k)U_2}{(c+k)U_3}}, \quad \text{and} \quad M = x^2 y^3 z^2.$$

Theorem 3.2, guarantee that we can find second first integral of the form

$$\Psi = x^{-1} y^{-2} z^{-1}.$$

Hence the first integrals of the required form are

$$H_1 = \phi^{\frac{U_3}{U_2}} \Psi^{-\frac{1}{2}\frac{U_2-U_3}{U_2}} = xy(1 + O(x,y,z)), \text{ and } H_2 = \phi^{-\frac{U_3}{U_2}} \Psi^{-\frac{1}{2}\frac{U_2+U_3}{U_2}} = yz(1 + O(x,y,z)).$$

Since $\zeta = xz\ell^{-1}$ satisfies $\dot\zeta = 2\zeta$, then the system (4.28) must be linearizable by Theorem 3.4.2.

For the case $U_2 = 0$, also $\ell = 1 + \frac{1}{2}(c+k)xz = 0$, is invariant algebraic surface with cofactor $L_\ell = (c+k)xz$ and an exponential factor $E_1 = e^{\frac{1}{2}U_3 y^2}$ with cofactor $L_{E_1} = -U_3 y^2$. One can easily find two first integrals of the form

$$H_1 = xy\ell^{-\frac{c}{c+k}} E_1^{-1} = xy(1 + O(x,y,z)), \text{ and } H_2 = yz\ell^{-\frac{k}{c+k}} E_1 = yz(1 + O(x,y,z)),$$

and it is also linearizable. Moreover, Since $\zeta = y$ satisfies $\dot\zeta = \zeta$, then the reduce system with $U_2 = 0$ must be linearizable by Theorem 3.4.2.

If $c = -k$, then we can find an exponential factor $E_2 = e^{\frac{1}{2}kxz - \frac{1}{2}U_3 y^2}$ with cofactor $L_{E_2} = kxz + U_3 y^2$. Hence the two first integrals of the reduce system are

$$H_1 = xyE_2 = xy \text{ and } H_2 = yzE_2^{-1}.$$

Since $\zeta = xz$ satisfies $\dot\zeta = 2\zeta$, then the reduce system must be linearizable by Theorem 3.4.2.

**Case 8. A)** if $(c+k)U_2 \neq 0$ the conditions reduce to

$$a - g = b - h = d + g = e + h = (c-k)U_2 + (c+k)U_3 - 2fU_2 =$$
$$2L_1 U_2(c-k) + U_3(c+k) - L_2 U_2(c+k) = L_3(c+k) + L_1(c-3k) =$$
$$2U_2(c-k) + U_3(c+k) - U_1(c+k) = V_1(c+k) - V_3(3c-k) = U_2 V_2 - U_3 V_3.$$

The system is

$$\dot{x} = x\left(1 + gxy + hyz + cxz + \frac{2U_2(c-k)+U_3(c+k)}{c+k}y^2 + \frac{V_3(3c-k)}{c+k}z^2\right),$$
$$\dot{y} = y\left(-1 - gxy - hyz + \frac{1}{2}\frac{U_2(c-k)+U_3(c+k)}{U_2}xz + \frac{L_1(2U_2(c-k)+U_3(c+k))}{(c+k)U_2}x^2 + U_2 y^2 + \frac{V_3 U_3}{U_2}z^2\right),$$
$$\dot{z} = z\left(1 + gxy + hyz + kxz - \frac{L_1(c-3k)}{c+k}x^2 + U_3 y^2 + V_3 z^2\right).$$

The invariant algebraic surface in this case is $\ell = 1 - \frac{2(c+k)gU_2}{U_2(c-3k)+U_3(c+k)}xy + \frac{2hU_2}{U_2-U_3}yz + \frac{1}{2}(c+k)xz + L_1 x^2 - U_2 y^2 + V_3 z^2 = 0$, with cofactor $L_\ell = (c+k)xz + 2L_1 x^2 + 2U_2 y^2 + 2V_3 z^2$. This gives two independent first integrals of the form

$$H_1 = xy\ell^{-\frac{1}{2}\frac{U_2(3c-k)+U_3(c+k)}{(c+k)U_2}}, \quad \text{and} \quad H_2 = yz\ell^{-\frac{1}{2}\frac{U_2+U_3}{U_2}}.$$

**B)** If $(c+k)U_2 = 0$, we consider several cases as follows
  1. If $U_2 = (c+k) = 0$, this is a sub-case of Case 7.
  2. If $U_2 = 0$ and $c+k \neq 0$, in this case we have $a+d = b+e = d+g = e+h = U_1 = U_2 = U_3 = 0$, this case separate for two cases

**References**

[3] Antonio Algaba, Cristóbal Garca, and Manuel Reyes. The analytic integrability problem for perturbations of homogeneous quadratic lotka-volterra systems. *arXiv preprint arXiv:1805.02873*, 2018.

[4] Waleed Aziz and Colin Christopher. Local integrability and linearizability of three-dimensional lotka–volterra systems. *Applied Mathematics and Computation*, 219(8):4067–4081, 2012.

[5] Waleed Aziz. Integrability and linearizability of three dimensional vector fields. *Qualitative theory of dynamical systems*, 13(2):197–213, 2014.

[6] Waleed Aziz. Integrability and linearizability problems of three dimensional

lotka–volterra equations of rank-2. *Qualitative Theory of Dynamical Systems*, pages 1–22, 2019.

[7] Faina Berezovskaya, Georgy Karev, and Roger Arditi. Parametric analysis of the ratio-dependent predator–prey model. *Journal of Mathematical Biology*, 43(3):221–246, 2001.

[8] Samuel Bottani and Basile Grammaticos. A simple model of genetic oscillations through regulated degradation. *Chaos, Solitons & Fractals*, 38(5):1468–1482, 2008.

[9] TC Bountis, A Ramani, B Grammaticos, and B Dorizzi. On the complete and partial integrability of non-hamiltonian systems. *Physica A: Statistical Mechanics and its Applications*, 128(1-2):268–288, 1984.

[10] Laurent Cairó and Jaume Llibre. Darboux integrability for 3d lotka-volterra systems. *Journal of Physics A: Mathematical and General*, 33(12):2395, 2000a.

[11] Laurent Cairó. Darboux first integral conditions and integrability of the 3d lotka-volterra system. *Journal of Nonlinear Mathematical Physics*, 7(4):511–531, 2000.

[12] J Chavarriga and J Giné. Integrability of cubic systems with degenerate infinity. *Differential Equations and Dynamical Systems*, 6(4):425–438, 1998.

[13] Xingwu Chen, Jaume Giné, Valery G Romanovski, and Douglas S Shafer. The 1:- q resonant center problem for certain cubic lotka–volterra systems. *Applied Mathematics and Computation*, 218(23):11620–11633, 2012.

[14] Yiannis T Christodoulides and Pantelis A Damianou. Darboux polynomials for lotka–volterra systems in three dimensions. *Journal of Nonlinear Mathematical Physics*, 16(3):339–354, 2009.

[15] Colin Christopher and Chengzhi Li. *Limit cycles of differential equations*. Springer Science & Business Media, 2007.

[16] Colin Christopher and Christiane Rousseau. Normalizable, integrable and linearizable saddle points in the lotka-volterra system. *Qualitative Theory of Dynamical Systems*, 5(1):11–61, 2004.

[17] C Christopher, Pavao Mardešic′, and C Rousseau. Normalizable, integrable, and linearizable saddle points for complex quadratic systems in $\mathbb{C}^2$. *Journal of Dynamical and Control Systems*, 9(3):311–363, 2003.

[18] C Christopher, Pavao Mardešic′, and C Rousseau. Normalizability, synchronicity, and relative exactness for vector fields in $\mathbb{C}^2$. *Journal of dynamical and control systems*, 10(4):501–525, 2004.

[19] Gaston Darboux. Mémoire sur les équations différentielles algébriques du premier ordre et du premier degré. *Bulletin des Sciences Mathématiques et*

*Astronomiques*, 2(1):151–200, 1878.

[20] Wolfram Decker. Singular 3-1-2-a computer algebra system for polynomial computations. *http://www. singular. uni-kl. de*, 2012.

[21] Freddy Dumortier, Jaume Llibre, and Joan C Artés. *Qualitative theory of planar differential systems*. Springer, 2006.

[22] Brigita Fercec, Xingwu Chen, and Valery G Romanovski. Integrability conditions for complex systems with homogeneous quintic nonlinearities. *J. Appl. Anal. Comput*, 1(1):9–20, 2011.

[23] Brigita Ferčec, Jaume Giné, Yirong Liu, and Valery G Romanovski. Integrability conditions for lotka-volterra planar complex quartic systems having homogeneous nonlinearities. *Acta applicandae mathematicae*, 124(1):107–122, 2013.

[24] Brigita Fercec and Jaume Giné. A blow-up method to prove formal integrability for some planar differential systems. *Journal Of Applied Analysis And Computation, 2018, vol. 8, núm. 6, p. 1833-1850*, 2018.

[25] Alexandra Fronville, Anton Sadovski, and Henryk Żołądek. Solution of the 1:- 2 resonant center problem in the quadratic case. *Fundamenta Mathematicae*, 157(2-3):191–207, 1998.

[26] Jaume Giné, Zhibek Kadyrsizova, Yirong Liu, and Valery G Romanovski. Linearizability conditions for lotka–volterra planar complex quartic systems having homogeneous nonlinearities. *Computers & Mathematics with Applications*, 61(4):1190–1201, 2011.

[27] Jaume Giné and Valery G Romanovski. Linearizability conditions for lotka–volterra planar complex cubic systems. *Journal of Physics A: Mathematical and Theoretical*, 42(22):225206, 2009.

[28] Jaume Giné and Valery G Romanovski. Integrability conditions for lotka–volterra planar complex quintic systems. *Nonlinear Analysis: Real World Applications*, 11(3):2100–2105, 2010.

[29] Alain Goriely. Integrability, partial integrability, and nonintegrability for systems of ordinary differential equations. *Journal of Mathematical Physics*, 37(4):1871–1893, 1996.

[30] Bidisha Goswami and Snehanshu Saha. Resource allocation modeling in abstraction using predator-prey dynamics: A qualitative analysis. *International Journal of Computer Applications*, 61(6), 2013.

[31] Bidisha Goswami, Jyotirmoy Sarkar, Snehanshu Saha, Saibal Kar, and Poulami Sarkar. Alvec: Auto-scaling by lotka volterra elastic cloud: A qos aware non linear dynamical allocation model. *Simulation Modelling Practice and Theory*, 93:262–292, 2019.

[32] Simon Gravel and Pierre Thibault. Integrability and linearizability of the lotka–volterra system with a saddle point with rational hyperbolicity ratio. *Journal of Differential Equations*, 184(1):20–47, 2002.

[33] Maoan Han, Valery G Romanovski, and Xiang Zhang. Integrability of a family of 2-dim cubic systems with degenerate infinity. *Rom. Journ. Phys*, 61:1–2, 2016.

[34] Zhaoping Hu, Maoan Han, and Valery G Romanovski. Local integrability of a family of three-dimensional quadratic systems. *Physica D: Nonlinear Phenomena*, 265:78–86, 2013.

[35] Zhaoping Hu, Valery G Romanovski, and Douglas S Shafer. 1:- 3 resonant centers on $\mathbb{C}^2$ with homogeneous cubic nonlinearities. *Computers & Mathematics with Applications*, 56(8):1927–1940, 2008.

[36] Yakovenko Ilyashenko. *Lectures on analytic differential equations*, volume 86. American Mathematical Soc., 2008.

[37] Vlastimil Křivan. On the gause predator–prey model with a refuge: A fresh look at the history. *Journal of Theoretical Biology*, 274(1):67–73, 2011.

[38] Willis E Lamb Jr. Theory of an optical maser. *Physical Review*, 134(6A):A1429, 1964.

[39] Changjian Liu, Guoting Chen, and Chengzhi Li. Integrability and linearizability of the lotka–volterra systems. *Journal of Differential Equations*, 198(2):301–320, 2004.

[40] Jaume Llibre and Carlos Gutiérrez. Darbouxian integrability for polynomial vector fields on the 2-dimensional sphere. *extracta mathematicae*, 17(2):289–301, 2002.

[41] Jaume Llibre and Clàudia Valls. Polynomial, rational and analytic first integrals for a family of 3-dimensional lotka-volterra systems. *Zeitschrift für angewandte Mathematik und Physik*, 62(5):761–777, 2011.

[42] Jaume Llibre. Integrability of polynomial differential systems. *Handbook of Differential Equations, Ordinary Differential Equations, Eds. A. Canada, P. Drabeck and A. Fonda, Elsevier*, pages 437–533, 2004.

[43] Jaume Llibre. On the integrability of the differential systems in dimension two and of the polynomial differential systems in arbitrary dimension. *Journal of Applied Analysis and Computation*, 1(1):0011–33, 2011.

[44] AJ Lotka. Elements of mathematical biology. ny, 1956.

[45] KE Malkin. Criteria for center of a differential equation. *Volg. Matem.*

*Sbornik*, 2:87–91, 1964.

[46] James D Meiss. *Differential dynamical systems*, volume 14. Siam, 2007.

[47] Adrian C Murza and Antonio E Teruel. Global dynamics of a family of 3-d lotka–volterra systems. *Dynamical Systems*, 25(2):269–284, 2010.

[48] VW Noonburg. A neural network modeled by an adaptive lotka-volterra system. *SIAM Journal on Applied Mathematics*, 49(6):1779–1792, 1989.

[49] Valery G Romanovski, Douglas S Shafer, et al. On the center problem for $p:-q$ resonant polynomial vector fields. *Bulletin of the Belgian Mathematical Society-Simon Stevin*, 15(5):871–887, 2008.

[50] Paul A Samuelson. A universal cycle? *Operations Researchverfahren*, 3:307, 1967.

[51] Vito Volterra. *Leçons sur la théorie mathématique de la lutte pour la vie*, volume 7. Gauthier-Villars, 1931.

[52] Qinlong Wang and Yirong Liu. Linearizability of the polynomial differential systems with a resonant singular point. *Bulletin des sciences mathematiques*, 132(2):97–111, 2008.

[53] Xiang Zhang. *Integrability of dynamical systems: algebra and analysis*, volume 47. Springer, 2017.

[54] Henryk Żołądek. The problem of center for resonant singular points of polynomial vector fields. *Journal of differential equations*, 137(1):94–118, 1997.